%% file: Paper.tex
\documentclass{llncs}

\usepackage[naustrian,english]{babel}
\usepackage{lmodern}
\usepackage[latin1]{inputenc}
\usepackage[T1]{fontenc}
\usepackage{url}
\usepackage[sans]{dsfont}
\usepackage{amssymb}
\usepackage{amsmath}
\usepackage{hyperref}
\usepackage{tikz}
\usetikzlibrary{positioning}
\usetikzlibrary{calc}
\usetikzlibrary{arrows}

\definecolor{col1}{RGB}{191,218,255}
\definecolor{col2}{RGB}{250,190,190}
\definecolor{col3}{RGB}{191,218,255}
\definecolor{col4}{RGB}{128,179,255}

\tikzset{thy/.style={
rectangle,
very thick,
fill=col1,
minimum height=5mm
}}

\newcommand{\Tma}{Theorema}
\newcommand{\Mma}{\textsl{Mathematica}}
\newcommand{\thy}[1]{\textsf{#1}}
\newcommand{\tma}[1]{\texttt{#1}}
\newcommand{\RR}{\mathcal{R}}
\newcommand{\TT}{\mathcal{T}}
\newcommand{\PP}{\mathcal{P}}
\newcommand{\TA}{\textsf{TheoryAnalyzer}}

\newcommand{\todo}[2][]{#2}
\newcommand{\comment}[1]{}

\title{Mathematical Theory Exploration in \Tma:\\Reduction Rings\thanks{This research was funded by the Austrian Science Fund (FWF): grant no. W1214-N15, project DK1}}
\author{Alexander Maletzky}
\institute{Doctoral Program ``Computational Mathematics'' and RISC\\Johannes Kepler University Linz, Austria\\[1mm]\href{mailto:alexander.maletzky@dk-compmath.jku.at}{\texttt{alexander.maletzky@dk-compmath.jku.at}}}

\begin{document}
\maketitle

\begin{abstract}
In this paper we present the first-ever computer formalization of the theory of Gr\"obner bases in reduction rings, which is an important theory in computational commutative algebra, in \Tma. Not only the formalization, but also the formal verification of all results has already been fully completed by now; this, in particular, includes the generic implementation and correctness proof of Buchberger's algorithm in reduction rings. Thanks to the seamless integration of proving and computing in \Tma, this implementation can now be used to compute Gr\"obner bases in various different domains directly within the system. Moreover, a substantial part of our formalization is made up solely by ``elementary theories'' such as sets, numbers and tuples that are themselves independent of reduction rings and may therefore be used as the foundations of future theory explorations in \Tma.

In addition, we also report on two general-purpose \Tma\ tools we developed for an efficient and convenient exploration of mathematical theories: an interactive proving strategy and a ``theory analyzer'' that already proved extremely useful when creating large structured knowledge bases.
\end{abstract}

\keywords{Gr\"obner bases, reduction rings, computer-supported theory exploration, automated reasoning, Theorema}

\section{Introduction}
\label{sec::Introduction}

This paper reports on the formalization and formal verification of the theory of reduction rings in \Tma\ that has recently been completed. Reduction rings, introduced by Buchberger in \cite{Buchberger1984}, generalize the domains where Gr\"obner bases can be defined and algorithmically computed from polynomial rings over fields to arbitrary commutative rings with identity, and may thus become more and more an important tool in computational commutative algebra, just as Gr\"obner bases in the original setting already are. Since definitions, theorems and proofs tend to be technical and lengthy, we are convinced that our formalization in a mathematical assistant system has the potential to facilitate the further development of the theory in the future (e.\,g. to non-commutative reduction rings).


To the best of our knowledge, reduction rings have never been the subject of formal theory exploration in \emph{any} software system so far; Gr\"obner bases in polynomial rings over fields have already been formalized in ACL2 \cite{Medina-Bulo2010}, Coq and OCaml \cite{Thery2001,Jorge2009} and Mizar \cite{Schwarzweller2006}, though, and a formalization in Isabelle by the author of this paper is currently in progress. Moreover, the purely algorithmic aspect (no theorems and proofs) of a variation of reduction rings was implemented in \Tma\ in \cite{Buchberger2003}. \Tma\ is also the software system we chose for our formalization, or, more precisely, \Tma~2.0 \cite{Windsteiger2014,Buchberger2016}.

The rest of this paper is organized as follows: Section~\ref{sec::ReductionRings} introduces the most important concepts of reduction rings and states the Main Theorem of the theory. Section~\ref{sec::Algorithm} presents Buchberger's algorithm for computing Gr\"obner bases in reduction rings as well as its implementation in \Tma, and briefly gives an idea about its correctness proof. Section~\ref{sec::Formalization} describes the overall formalization of the theory and its individual components in a bit more detail, and Section~\ref{sec::Tools} presents the interactive proving strategy and the \TA\ tool that we developed and already heavily used in the course of the formalization and that will be useful also in future theory explorations. Section~\ref{sec::Conclusion}, finally, summarizes our findings and contains an outlook on future work.

\input{ReductionRings}
\input{Algorithm}
\input{Formalization}
\input{Tools}

\section{Conclusion}
\label{sec::Conclusion}

The formal treatment of reduction ring theory and the various elementary mathematical theories as presented in this paper might not only serve as the basis of future theory explorations in \Tma, but already had positive effects on the theory \emph{itself}: during the verification, two minor problems in the literature on reduction rings were discovered and immediately fixed in our formalization. The first problem is related to the notion of \emph{irrelativity} as introduced in \cite{Stifter1989} and explained in more detail in \cite{Maletzky2015}. The second problem concerns fields as reduction rings: in an infinite field, two elements have \emph{infinitely many} minimal non-trivial common reducibles (mntcr's), although for an algorithmic treatment one axiom of reduction rings requires the number of mntcrs to be finite.\footnote{This problem was already known in \cite{Buchberger1984}, but no attempts have been made to fix it so far.} We solved this problem by introducing an equivalence relation in reduction rings and weakening the axiom to accept a finite number of \emph{equivalence classes} of mntcrs.

There are many possibilities for future work. On the theory level, other aspects of, and approaches to, Gr\"obner bases (again in the original setting) could be formalized, for instance the computation of Gr\"obner bases by matrix triangularizations \cite{Wiesinger-Widi2015}. For this, the further improvement of the tools described in Sect.~\ref{sec::Tools} and the development of new tools might be necessary (more flexible interactive proving strategy, proof checker, \ldots).

\paragraph*{Acknowledgments} I thank my doctoral adviser Bruno Buchberger and Wolfgang Windsteiger for many stimulating and inspiring discussions about Gr\"obner bases, formal mathematics and \Tma.\\[1mm]
This research was funded by the Austrian Science Fund (FWF): grant no. W1214-N15, project DK1

\bibliography{References}

\end{document}

%% file: ReductionRings.tex
\section{Gr\"obner Bases and Reduction Rings}
\label{sec::ReductionRings}

In this section we review the main concepts of the theory whose formal treatment in \Tma\ is the content of this paper. To this end, we first give a short motivation of Gr\"obner bases and reduction rings, and then present the most important definitions and results of the theory. A far more thorough introduction can be found in the literature, e.\,g. in \cite{Adams1994}.

Originally, the theory of Gr\"obner bases was invented for multivariate polynomial rings over fields. There, it can be employed to decide the ideal membership problem, to solve systems of algebraic equations, and many more, and hence is of great importance in computer algebra and many other areas of mathematics, computer science, engineering, etc.

Because of their ability to solve non-trivial, frequently occurring problems in mathematics, it is only natural to try to generalize Gr\"obner bases from polynomial rings over fields to other algebraic structures. And indeed, nowadays quite some generalizations exist: to non-commutative polynomial rings, to polynomial rings over the integers and other Euclidean- or integral domains, and many more. Reduction rings are a generalization as well, but in a slightly different spirit: in contrast to the other generalizations, reduction rings do not require the domain of discourse to have any polynomial structure. Instead, \emph{arbitrary} commutative rings with identity element may in principle be turned into reduction rings, only by endowing them with some additional structure (see below). It must be noted, however, that not \emph{every} commutative ring with identity can be made a reduction ring; known examples of reduction rings are all fields, the integers, quotient rings of integers modulo arbitrary $n\in\bbbn$ (which may contain 
zero-divisors!), and polynomial rings over reduction rings.

\subsection{Reduction Rings}
\label{sec::RR}

Reduction rings were first introduced by Buchberger in 1984 \cite{Buchberger1984} and later further generalized by Stifter in the late-1980s \cite{Stifter1988,Stifter1989}; our formalization is mainly based on \cite{Stifter1989}. Here, we only recall the key ideas and main definitions and results of the theory. For this, let in the sequel $\RR$ be a commutative ring with identity (possibly containing zero-divisors).

In order to turn $\RR$ into a reduction ring, it first and foremost has to be endowed by two additional entities: a function $M:\RR\rightarrow\mathcal{P}(\RR)$ that maps every ring element $c$ to a set of ring elements (denoted by $M_c$) called the \emph{set of multipliers} of $c$, and a partial Noetherian (i.\,e. well-founded) order relation $\preceq$. With these ingredients it is possible to introduce the crucial notion of reduction rings, namely that of \emph{reduction}:

\begin{definition}[Reduction]
Let $C\subseteq\RR$. The reduction relation modulo $C$, denoted by $\rightarrow_C$, is a binary relation on $\RR$ such that $a\rightarrow_C b$ iff $b \prec a$ and there exists some $c\in C$ and some $m\in M_c$ such that $b=a-m\,c$.\\
As usual, $\rightarrow_C^*$ and $\leftrightarrow_C^*$ denote the reflexive-transitive- and the symmetric-reflexive-transitive closure of $\rightarrow_C$, respectively. Moreover, for a given $z\in\RR$, $a$ and $b$ are said to be \emph{connectible below} $z$, denoted by $a\leftrightarrow_C^{\prec z} b$, iff $a\leftrightarrow_C^* b$ and all elements in the chain between $a$ and $b$ are strictly less than $z$ (w.\,r.\,t. $\preceq$).
\end{definition}

Of course, the function $M$ and the relation $\preceq$ cannot be chosen arbitrarily but, together with the usual ring operations, have to satisfy certain non-trivial constraints, the so-called \emph{reduction ring axioms}. In total, there are 14 of them, with some being quite simple ($0$ must be the least element w.\,r.\,t. $\preceq$, for instance), others are extremely technical. The complete list underlying our formalization is omitted here because of space limitations but can be found in \cite{Maletzky2015}.

Note that in reduction rings $\leftrightarrow_C^*$ coincides with the congruence relation modulo the ideal generated by $C$. Hence, if it is possible to decide $\leftrightarrow_C^*$, then the ideal membership problem could effectively be solved -- and this is where Gr\"obner bases come into play.

\subsection{Gr\"obner Bases}
\label{sec::GB}

We can start with the definition of Gr\"obner bases in reduction rings right away:
\begin{definition}[Gr\"obner basis]
Let $G\subseteq\RR$. Then $G$ is called a \emph{Gr\"obner basis} iff $G$ is finite and $\rightarrow_G$ is Church-Rosser, i.\,e. whenever $a\leftrightarrow_G^* b$ there exists a common successor $s$ with $a\rightarrow_G^* s$ and $b\rightarrow_G^* s$.\\
For $C\subseteq\RR$, $G$ is called a Gr\"obner basis of $C$ iff it is a Gr\"obner basis and $\langle G\rangle$ (i.\,e. the ideal generated by $G$ over $\RR$) is the same $\langle C\rangle$.
\end{definition}

If reduction can effectively be carried out, i.\,e. whenever $a$ is reducible modulo $C$ then some $b$ with $a\rightarrow_C b$ can be computed, and for any given $C\subseteq\RR$ a Gr\"obner basis $G$ of $C$ exists and can be computed, then the problem of deciding membership in $\langle C\rangle$ can be solved: a given candidate $a$ simply has to be totally reduced modulo $G$ until an irreducible element $h$ is obtained; then $a\in\langle C\rangle$ iff $h=0$.

The axioms of reduction rings ensure that for every $C\subseteq\RR$ a Gr\"obner basis does not only exist, but can even be effectively computed (see Section~\ref{sec::Algorithm}). This key result is based on the following

\begin{theorem}[Buchberger's Criterion]
\label{thm::Main}
Let $G\subseteq\RR$ finite. Then $G$ is a Gr\"obner basis iff for all $g_1,g_2\in G$ and all \emph{minimal non-trivial common reducibles} $z$ of $g_1$ and $g_2$, $a_1\leftrightarrow_G^{\prec z} a_2$, where $z\rightarrow_{\{g_i\}} a_i$ for $i=1,2$ ($(a_1,a_2)$ is called a \emph{critical pair} of $g_1$ and $g_2$ w.\,r.\,t. $z$).
\end{theorem}
The precise definition of \emph{minimal non-trivial common reducible} (mntcr) is slightly technical and omitted here; the interested reader may find it in the referenced literature. Intuitively, a mntcr of $g_1$ and $g_2$ is an element that can be reduced both modulo $\{g_1\}$ and modulo $\{g_2\}$ in a \emph{non-trivial} way\footnote{In polynomial rings over fields, the mntcr of two polynomials is just the least common multiple of the leading terms of the polynomials.}


%% file: Algorithm.tex
\section{Buchberger's Algorithm}
\label{sec::Algorithm}

\begin{figure}[t]
\centering
\includegraphics[width=\linewidth]{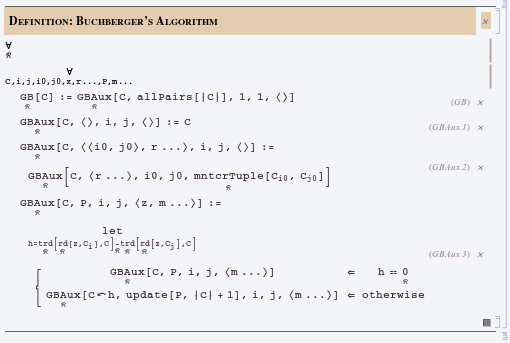}
\caption{Buchberger's algorithm in \Tma.}
\label{fig::Algorithm}
\end{figure}

Theorem~\ref{thm::Main} not only contains a finite criterion for checking whether a given set $G$ is a Gr\"obner basis or not, but it even gives rise to an algorithm for actually \emph{computing} Gr\"obner bases. This algorithm, presented in Fig.~\ref{fig::Algorithm}, is a critical-pair/completion algorithm that, given an input set $C\subseteq\RR$, basically checks the criterion of Thm.~\ref{thm::Main} for all pairs of elements of $C$, and if it fails for a pair $(C_i,C_j)$, then $C$ is \emph{completed} by a new element $h$ that makes the criterion hold for $(C_i,C_j)$. Of course, afterward all pairs involving the new element $h$ have to be considered as well.

Figure~\ref{fig::Algorithm} presents the algorithm as implemented in a functional style in \Tma. Function \tma{GB} is the main function that takes as input the tuple\footnote{\tma{GB} is implemented for tuples rather than sets, for practical reasons.} $C$ a Gr\"obner basis shall be computed for. It then calls \tma{GBAux} with suitable initial arguments, whose first argument serves as the accumulator of the tail-recursive function. Its second argument is the tuple of all pairs of indices of $C$ that have not been dealt with yet, and its third and fourth arguments are the indices $i$ and $j$ of the elements currently under consideration. The last argument, finally, is the tuple of all mntcrs of $C_i$ and $C_j$ that still have to be checked. Formula (GBAux 3) is the crucial one: The constituents of the critical pair originating from $C_i$ and $C_j$ and mntcr $z$ are totally reduced modulo the current basis $C$, and the difference is assigned to $h$. If $h=0$, the critical pair can be connected below $z$ 
according to the condition in Thm.~\ref{thm::Main}, so nothing else has to be done in this case. Otherwise, $h$ is added to $C$, ensuring connectibility below the new basis, and the index-pair-tuple is updated to include also the pairs involving the new element $h$.

Buchberger's algorithm, or, more precisely, function \tma{GB}, can be proved to behave according to the following specification:
\begin{quote}
If $\RR$ is a reduction ring and $C$ is a tuple of elements of $\RR$, \tma{GB} terminates and returns again a tuple $G$ of elements of $\RR$. $G$ is a Gr\"obner basis of $C$.
\end{quote}
The proof of this claim was carried out formally in \Tma. It heavily depends on Thm.~\ref{thm::Main}, of course, but also quite some other technicalities (concerning the indices, for instance) have to be taken into account. Furthermore, termination of \tma{GBAux} is by no means obvious: its second argument, which must eventually become empty, is enlarged in the second case of (GBAux 3), meaning that this case must be shown to occur only finitely often. A separate reduction ring axiom is needed to ensure this.

Function \tma{GB} is not only of theoretical interest for our formalization, but can also be executed on concrete input to actually compute Gr\"obner bases, provided that the underlying domain $\RR$ is a reduction ring and implements a couple of auxiliary functions \tma{GB} depends upon (most importantly, the usual ring operations). At the moment, the following domains included in the formalization meet these requirements; the proofs thereof are part of the formalization, of course (see also Sect.~\ref{sec::RRT}):

\begin{itemize}
 \item all fields, in particular the \Tma\ built-in fields $\bbbq$, $\bbbr$ and $\bbbc$,
 \item $\bbbz$,
 \item $\bbbz_n=\bbbz/n\bbbz$ for arbitrary $n\in\bbbn$,
 \item multivariate polynomial rings over the aforementioned domains.
\end{itemize}

\begin{figure}[t]
\centering
\includegraphics[width=0.65 \linewidth]{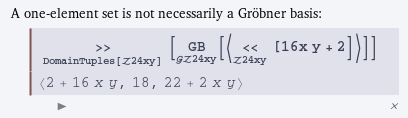}
\caption{A sample computation in \Tma. The ``\tma{<\,<}'' and ``\tma{>\,>}'' are only responsible for the in- and output of polynomials and do not affect the actual computation.}
\label{fig::Computation}
\end{figure}

Function \tma{GB} always returns provenly correct results when used in these domains. Figure~\ref{fig::Computation} shows a sample computation in $\bbbz_{24}[x,y]$, carried out directly within \Tma.

For the sake of completeness we have to point out that Buchberger's algorithm and Thm.~\ref{thm::Main} as presented here were simplified a bit compared to our actual formalization. For one thing, the sets of multipliers $M_c$ have to be split into several (finitely many) indexed subsets $M_c^i$, and the notion of mntcr depends on these indices; mntcrs for \emph{all} pairs of indices have to be considered separately, both in the theorem and in the algorithm. Also, the actual implementation of \tma{GB} employs the so-called \emph{chain criterion} for avoiding useless reductions; this criterion, hence, increases efficiency and works in reduction rings in pretty much the same way as in the original setting of polynomials over fields, see \cite{Buchberger1979}. The interested reader is referred to \cite{Maletzky2015} for an unsimplified statement of Thm.~\ref{thm::Main}, and to \cite{Maletzky2015a} for a more detailed discussion of Buchberger's algorithm in our formalization.

%% file: Formalization.tex
\section{Structure of the Formalization}
\label{sec::Formalization}

In this section we have a closer look at the formalization of all of reduction ring theory in \Tma. In particular, the emphasis is on how the theory is split into smaller sub-theories, what these sub-theories consist of, how they are related to each other, and how big they are in terms of formulas and proofs.

Although the paper has only been about reduction rings so far, it must be noted that a substantial part of our formalization is actually concerned with rather basic concepts, such as sets, algebraic structures, numbers, tuples (or lists) and sequences that are themselves independent of reduction ring theory and merely serve as its logical backbone. In this respect, our formalization can also be regarded a major contribution to a structured knowledge base of elementary mathematical theories in \Tma~2.0 that can be reused in future theory explorations. Such a knowledge base did not exist in \Tma~2.0 before, which justifies, in our opinion, presenting it just alongside the formal treatment of reduction rings in this section (only superficially, though).

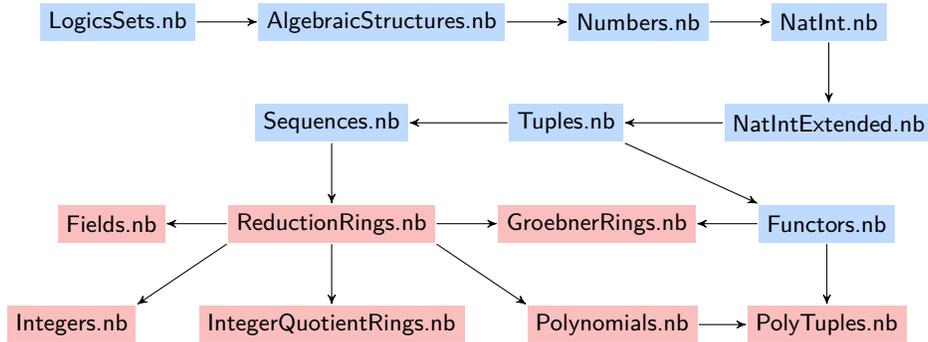
\begin{figure}[t]
\centering
\begin{tikzpicture}[text centered, node distance=8mm and 8mm]
\node (ls) [thy] {\thy{LogicsSets.nb}};
\node (as) [thy, right=of ls] {\thy{AlgebraicStructures.nb}};
\node (nu) [thy, right=of as] {\thy{Numbers.nb}};
\node (ni) [thy, right=of nu] {\thy{NatInt.nb}};
\node (ne) [thy, below=of ni] {\thy{NatIntExtended.nb}};
\node (tu) [thy, left=of ne, xshift=-5mm] {\thy{Tuples.nb}};
\node (sq) [thy, left=of tu, xshift=-5mm] {\thy{Sequences.nb}};
\node (rr) [thy, fill=col2, below=of sq] {\thy{ReductionRings.nb}};
\node (gr) [thy, fill=col2, right=of rr] {\thy{GroebnerRings.nb}};
\node (fn) [thy, right=of gr] {\thy{Functors.nb}};
\node (fl) [thy, fill=col2, left=of rr] {\thy{Fields.nb}};
\node (iq) [thy, fill=col2, below=of rr] {\thy{IntegerQuotientRings.nb}};
\node (in) [thy, fill=col2, left=of iq] {\thy{Integers.nb}};
\node (py) [thy, fill=col2, right=of iq] {\thy{Polynomials.nb}};
\node (pt) [thy, fill=col2, below=of fn] {\thy{PolyTuples.nb}};
\pgfsetarrowsend{stealth'}
\draw (ls.east) -- (as.west);
\draw (as.east) -- (nu.west);
\draw (nu.east) -- (ni.west);
\draw (ni.south) -- (ne.north);
\draw (ne.west) -- (tu.east);
\draw (tu.south east) -- (fn.north west);
\draw (tu.west) -- (sq.east);
\draw (sq.south) -- (rr.north);
\draw (rr.west) -- (fl.east);
\draw (rr.south west) -- (in.north east);
\draw (rr.south) -- (iq.north);
\draw (rr.south east) -- (py.north west);
\draw (rr.east) -- (gr.west);
\draw (py.east) -- (pt.west);
\draw (fn.west) -- (gr.east);
\draw (fn.south) -- (pt.north);
\end{tikzpicture}
\caption{The theory dependency graph.}
\label{fig::ThyGraph}
\end{figure}

Figure~\ref{fig::ThyGraph} shows the dependencies of the individual sub-theories on each other. Each node represents a sub-theory, contained in a separate \Tma\ notebook, and a directed edge from theory $A$ to theory $B$ means that $B$ logically depends on $A$ in the sense that formulas (i.\,e. definitions or theorems) contained in $A$ were used in the proof of a theorem in $B$. The color of a node indicates whether the corresponding theory belongs to the knowledge base of elementary theories (\todo{blue}; Sect.~\ref{sec::ET}) or is directly related to reduction rings (\todo{red}; Sect.~\ref{sec::RRT}). Note also that transitive edges are omitted for better readability, e.\,g. theory \thy{Numbers.nb} not only depends \emph{indirectly} on theory \thy{LogicSets.nb} (via \thy{AlgebraicStructures.nb}), but also \emph{directly}; this fact is not reflected in Fig.~\ref{fig::ThyGraph}.

\begin{figure}[t]
\centering
\begin{tikzpicture}[text centered, node distance=4mm and 2mm]
\node (ls) [thy, fill=none] {\thy{LogicSets.nb}};
\node (as) [thy, fill=none, below=of ls.east, anchor=east] {\thy{AlgebraicStructures.nb}};
\node (nu) [thy, fill=none, below=of as.east, anchor=east] {\thy{Numbers.nb}};
\node (ni) [thy, fill=none, below=of nu.east, anchor=east] {\thy{NatInt.nb}};
\node (ne) [thy, fill=none, below=of ni.east, anchor=east] {\thy{NatIntExtended.nb}};
\node (tu) [thy, fill=none, below=of ne.east, anchor=east] {\thy{Tuples.nb}};
\node (sq) [thy, fill=none, below=of tu.east, anchor=east] {\thy{Sequences.nb}};
\node (fn) [thy, fill=none, below=of sq.east, anchor=east] {\thy{Functors.nb}};
\node (rr) [thy, fill=none, below=of fn.east, anchor=east] {\thy{ReductionRings.nb}};
\node (fl) [thy, fill=none, below=of rr.east, anchor=east] {\thy{Fields.nb}};
\node (in) [thy, fill=none, below=of fl.east, anchor=east] {\thy{Integers.nb}};
\node (iq) [thy, fill=none, below=of in.east, anchor=east] {\thy{IntegerQuotientRings.nb}};
\node (py) [thy, fill=none, below=of iq.east, anchor=east] {\thy{Polynomials.nb}};
\node (pt) [thy, fill=none, below=of py.east, anchor=east] {\thy{PolyTuples.nb}};
\node (gr) [thy, fill=none, below=of pt.east, anchor=east] {\thy{GroebnerRings.nb}};

\node[fill=col3,text width=51.4mm,right=of ls,align=right,inner sep=2pt] {\scriptsize 54};
\node[fill=col4,text width=40.6mm,right=of ls,align=right,inner sep=2pt] {\scriptsize 203};

\node[fill=col3,text width=19.8mm,right=of as,align=right,inner sep=2pt] {\scriptsize 39};
\node[fill=col4,text width=12mm,right=of as,align=right,inner sep=2pt] {\scriptsize 60};

\node[fill=col3,text width=41.4mm,right=of nu,align=right,inner sep=2pt] {\scriptsize 28};
\node[fill=col4,text width=35.8mm,right=of nu,align=right,inner sep=2pt] {\scriptsize 179};

\node[fill=col3,text width=65.8mm,right=of ni,align=right,inner sep=2pt] {\scriptsize 29};
\node[fill=col4,text width=60mm,right=of ni,align=right,inner sep=2pt] {\scriptsize 300};

\node[fill=col3,text width=53.6mm,right=of ne,align=right,inner sep=2pt] {\scriptsize 24};
\node[fill=col4,text width=48.8mm,right=of ne,align=right,inner sep=2pt] {\scriptsize 244};

\node[fill=col3,text width=60mm,right=of tu,align=right,inner sep=2pt] {\scriptsize 29};
\node[fill=col4,text width=54.2mm,right=of tu,align=right,inner sep=2pt] {\scriptsize 271};

\node (sqf) [fill=col3,text width=11mm,right=of sq,text=col3,inner sep=2pt] {\scriptsize 0};
\node[right] at (sqf.east) {\scriptsize 7};
\node[fill=col4,text width=9.6mm,right=of sq,align=right,inner sep=2pt] {\scriptsize 48};

\node (fnf) [fill=col3,text width=9.0mm,right=of fn,align=right,text=col3,inner sep=2pt] {\scriptsize 0};
\node[right] at (fnf.east) {\scriptsize 9};
\node[fill=col4,text width=7.2mm,right=of fn,align=right,inner sep=2pt] {\scriptsize 36};

\node[fill=col3,text width=62.4mm,right=of rr,align=right,inner sep=2pt] {\scriptsize 60};
\node[fill=col4,text width=50.4mm,right=of rr,align=right,inner sep=2pt] {\scriptsize 252};

\node (flf) [fill=col3,text width=10.4mm,right=of fl,align=right, text=col3,inner sep=2pt] {\scriptsize 0};
\node[right] at (flf.east) {\scriptsize 16};
\node[fill=col4,text width=7.2mm,right=of fl,align=right,inner sep=2pt] {\scriptsize 36};

\node (inf) [fill=col3,text width=20.4mm,right=of in,align=right, text=col3,inner sep=2pt] {\scriptsize 0};
\node[right] at (inf.east) {\scriptsize 17};
\node[fill=col4,text width=17mm,right=of in,align=right,inner sep=2pt] {\scriptsize 85};

\node (iqf) [fill=col3,text width=19mm,right=of iq,align=right,text=col3,inner sep=2pt] {\scriptsize 0};
\node[right] at (iqf.east) {\scriptsize 17};
\node[fill=col4,text width=15.6mm,right=of iq,align=right,inner sep=2pt] {\scriptsize 78};

\node[fill=col3,text width=77.8mm,right=of py,align=right,inner sep=2pt] {\scriptsize 48};
\node[fill=col4,text width=68.2mm,right=of py,align=right,inner sep=2pt] {\scriptsize 341};

\node[fill=col3,text width=43.6mm,right=of pt,align=right,inner sep=2pt] {\scriptsize 57};
\node[fill=col4,text width=32.2mm,right=of pt,align=right,inner sep=2pt] {\scriptsize 161};

\node[fill=col3,text width=44.0mm,right=of gr,align=right,inner sep=2pt] {\scriptsize 50};
\node[fill=col4,text width=34.0mm,right=of gr,align=right,inner sep=2pt] {\scriptsize 170};

\end{tikzpicture}
\caption{The sizes of the individual sub-theories. The larger number in each row corresponds to the proved theorems, the smaller one to the definitions and axioms.}
\label{fig::ThyStats}
\end{figure}
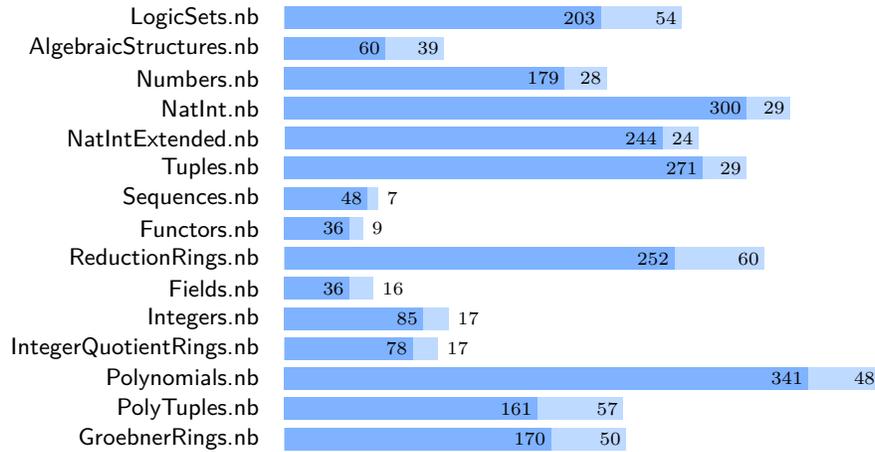

Figure~\ref{fig::ThyStats} displays the sizes of the individual sub-theories in terms of the numbers of proved and unproved formulas. The total number of proved theorems in the whole formalization in 2464, the total number of unproved definitions and axioms is 484. Hence, the total number of formulas is \textbf{2948}.

At the moment, the formalization with all \Tma\ notebooks and proofs is not yet publicly available (e.\,g. in an online repository), because the mechanism for turning \Tma\ theories into so-called \emph{\Tma\ Knowledge Archives} that can easily be shared amongst the users of the system is still in the development stages. As soon as it is completed, we will immediately put our formalization into a public repository that will be linked on the official \Tma\ web page.\footnote{\url{http://www.risc.jku.at/research/theorema/software/}} The interested reader may nevertheless obtain the full formalization (or part of it) in its current form by contacting the author.

\subsection{Elementary Theories}
\label{sec::ET}

Most of the sub-theories in this category have rather self-explanatory names, and we will not go into details regarding their contents. Some remarks are still in place, though.

Theories \thy{Numbers.nb}, \thy{NatInt.nb} and \thy{NatIntExtended.nb} are all about natural numbers and integers: the very definition of natural numbers by purely set-theoretic means, as well as the definition of integers as some quotient domain of pairs of natural numbers are contained in \thy{Numbers.nb}, and the other two theories basically consist of hundreds of results about linear and non-linear arithmetic, division with quotient and remainder, the greatest common divisor, finite sums and mappings from $\bbbn$ to $\bbbn$ (needed for infinite sequences).

Theory \thy{Functors.nb} contains a couple of general \Tma\ functors, mainly for constructing product domains from given ones.\footnote{For information on functors and domains in \Tma, see \cite{Windsteiger1999,Buchberger2003}} The most important functor in this theory, \tma{LexOrder}, maps two ordered domains to their lexicographic product; this functor was needed for proving termination of function \tma{GB} (see Sect.~\ref{sec::Algorithm}). \thy{Functors.nb} also proves that the order in the new domain is still partial/total/Noetherian if the orders in the original domains are.

In general one must note that the elementary theories so far only include mathematical content that was explicitly needed for the formal treatment of reduction ring theory. Although this is quite comprehensive and covers many notions and concepts, it is still fairly incomplete.

\subsection{Reduction Ring Theory}
\label{sec::RRT}

\paragraph{\thy{ReductionRings.nb}}
\label{thy::RR}

contains the definitions of several auxiliary notions in reduction rings, like reducibility, the reduction relation (and its various closures) and properties of binary relations (confluence, local confluence, Church-Rosser), as well as the definitions of reduction rings and Gr\"obner bases. Reduction rings are defined through a unary predicate, \tma{isReductionRing}, that is simply the conjunction of all reduction ring axioms together with the axioms of commutative rings with identity.

Besides these definitions, the main contents of \thy{ReductionRings.nb} are the Main Theorem of reduction ring theory, Thm.~\ref{thm::Main}, and the theorem that states that the symmetric-reflexive-transitive closure of the reduction relation modulo a set $C$ coincides with ideal congruence modulo the same set $C$, together with their proofs. The proof of Thm.~\ref{thm::Main} is non-trivial and lengthy, which is reflected by the fact that many auxiliary lemmas were needed before it could finally be completed, and one of these lemmas in fact deserves special attention: the \emph{Generalized Newman Lemma}. The Generalized Newman Lemma is a general result about sufficient conditions for binary relations to be confluent (and thus Church-Rosser) that was first introduced in \cite{Winkler1983}.

Please note that everything in this theory is \emph{non-algorithmic} in the sense that no single algorithm is implemented or specified. All algorithmic aspects of our formal reduction ring theory, in particular Buchberger's algorithm for computing Gr\"obner bases, are part of \thy{GroebnerRings.nb}.

\paragraph{\thy{GroebnerRings.nb}}
\label{thy::GR}

contains all the algorithmic aspects of the formalization, like the implementation and specification of Buchberger's algorithm. More precisely, the theory contains a functor called \tma{GroebnerRing} that extends a given input domain $D$ by the function \tma{GB} that implements Buchberger's algorithm and can thus be used for computing Gr\"obner bases. \tma{GB} is defined in terms of auxiliary functions provided by the underlying domain $D$, such as the basic ring operations and the partial Noetherian ordering in reduction rings. However, following a general principle of functors and domains in \Tma, $D$ can be completely arbitrary: it does not need to be a reduction ring, nor even a ring, meaning that some operations used in function \tma{GB} are possibly undefined -- and this is perfectly fine, except that one cannot expect to obtain a Gr\"obner basis when calling the function. But if $D$ \emph{is} a reduction ring, i.\,e. \tma{isReductionRing[$D$]} holds, then the function really behaves according to its specification. The proof of this claim is non-trivial, even if Thm.~\ref{thm::Main} is already known, and also contained in \thy{GroebnerRings.nb}.

In addition to the implementation, specification and correctness proof of Buchberger's algorithm, various sample computations of Gr\"obner bases in different domains ($\bbbz_{24}$, $\bbbz_{24}[x,y]$, $\bbbq[x,y,z]$, for instance) are included in \thy{GroebnerRings.nb} as well.

\paragraph{\thy{Fields.nb}}
\label{thy::Fl}

contains a \Tma\ functor, \tma{ReductionField}, that takes an input domain $K$ and extends it by those objects (function $M$ and relation $\preceq$) that turn $K$ into a reduction ring. These new objects are defined in such a way that if $K$ is a field, then the extension really \emph{is} a reduction ring -- otherwise nothing can be said about it. The proof of this claim is of course also contained in \thy{Fields.nb}, and actually it is quite straight-forward, as can be seen from Fig.~\ref{fig::ThyStats}.

\paragraph{\thy{Integers.nb}}
\label{thy::In}

contains a \Tma\ functor, \tma{ReductionIntegers}, that does not take any input domains but simply constructs a new domain whose carrier is $\bbbz$ and that provides the additional objects for turning $\bbbz$ into a reduction ring, following \cite{Buchberger1984}. The proof of this claim is included in the theory as well.

\paragraph{\thy{IntegerQuotientRings.nb}}
\label{thy::IQ}

contains a \Tma\ functor, \tma{ReductionIQR}, that takes a positive integer $n$ and constructs a new domain whose carrier is the set $\{0,\ldots,n-1\}$ and that provides the additional objects for turning $\bbbz_n$, represented by $\{0,\ldots,n-1\}$, into a reduction ring, following \cite{Stifter1988}. The proof of this claim is of course included in the theory as well. Surprisingly, although turning $\bbbz_n$ into a reduction ring is more involved than $\bbbz$\footnote{The first attempt in \cite{Buchberger1984} was erroneous.}, fewer auxiliary results were needed in \thy{IntegerQuotientRings.nb} than in \thy{Integers.nb} (see Fig.~\ref{fig::ThyStats}). This is due to the fact that the reduction ring ordering $\preceq$ in $\bbbz_n$ is much simpler than in $\bbbz$.

\paragraph{\thy{Polynomials.nb}}
\label{thy::Po}

contains the general result that the $n$-variate polynomial ring over a reduction ring is again a reduction ring, if the sets of multipliers and the order relation are defined appropriately. This is accomplished by first introducing the class of \emph{reduction polynomial domains} over a coefficient domain $\RR$ and a power-product domain $\TT$. A domain $\PP$ belongs to this class iff it provides the usual ring operations, a coefficient function that maps each power-product from $\TT$ to a coefficient in $\RR$, a set of multipliers for each element in $\PP$ (i.\,e. the function $M$), and an order relation $\preceq$, and all these objects satisfy certain constraints (e.\,g. the coefficient function must have finite support and must interact with $+$ and $\cdot$ in the usual way, the sets of multipliers must be of a particular form, and the ordering must be defined in a certain way). These constraints, whose precise formulations can be found in \cite{Buchberger1984}, ensure that if $\RR$ is a reduction ring and $\TT$ 
is a domain of commutative power-products, then $\PP$ is a reduction ring as well. This is one of the fundamental results of reduction ring theory, and its proof is very complicated and tedious (even more complicated than the proof of Thm.~\ref{thm::Main}, as can be seen from Fig.~\ref{fig::ThyStats}). Nevertheless, it has been entirely completed already and is also part of \thy{Polynomials.nb}.

Note that all definitions and results in this theory are on a very abstract level: no concrete representation of multivariate polynomials, be it as tuples of monomials, as iterated univariate polynomials, or whatsoever, is ever mentioned in the whole theory, but instead polynomials are essentially viewed as functions from $\TT$ to $\RR$ with finite support. This approach has the advantage that the results can easily be specialized to many \emph{different} representations of polynomials, if necessary, and this is just what is made use of in theory \thy{PolyTuples.nb}.

\paragraph{\thy{PolyTuples.nb}}
\label{thy::PT}

contains a functor, \tma{PolyTuples}, that takes two domains $\RR$ and $\TT$ as input and constructs the domain $\PP$ of reduction-polynomials over coefficient domain $\RR$ and power-product domain $\TT$ represented as ordered (w.\,r.\,t. the ordering on $\TT$) tuples of monomials. Monomials, in turn, are represented as pairs of coefficients and power-products. $\PP$ provides the additional functions and relations needed to prove that it belongs to the class of reduction polynomial domains, and thus is a reduction ring thanks to the key result in \thy{Polynomials.nb}.\footnote{Once again, this is only true if $\RR$ is a reduction ring and $\TT$ is a domain of commutative power-products.} The proof of this claim is part of the theory, of course.

Besides functor \tma{PolyTuples}, three additional functors for constructing domains of commutative power-products are also contained in \thy{PolyTuples.nb}: one for a purely lexicographic term order, one for a degree-lexicographic term order, and one for a degree-reverse-lexicographic term order (see, e.\,g., \cite{Robbiano1985a}). In either case, power-products are represented as tuples of natural numbers.

%% file: Tools.tex
\section{New Tools}
\label{sec::Tools}

In this section we present two useful tools that we developed in the course of the formalization of reduction rings: an interactive proving strategy and a mechanism for analyzing the logical structure of \Tma\ theories. As will be seen in the following two subsections, the tools are general-purpose tools and thus completely independent of our concrete formalization, and hence may be used in any other theory exploration in \Tma\ as well. For that reason, they are planned to be integrated into the official version of the system in the near future.

\subsection{Interactive Proving Strategy}
\label{sec::Interactive}

Originally, \Tma\ focused very much on \emph{automated} proving where the user only initiates a proof attempt and then waits until the system either fails or succeeds, without any possibility for interaction. It soon became clear, though, that this approach was too restrictive, and so a mechanism for doing proofs interactively was added to \Tma~1 in \cite{Piroi2005}. This also influenced the design of the new version of the system, \Tma~2.0, in that it by default provides two pre-defined possibilities for interactively guiding the proof search: instantiating quantified formulas and choosing the most promising among several alternative branches in the proof tree. However, after the completion of the formalization of the complexity analysis of Buchberger's algorithm in the bivariate case \cite{Maletzky2014}, we realized that this still was not enough for efficiently proving the long and complicated theorems that awaited us in the theory of reduction rings -- and this, finally, triggered the implementation of a \emph{fully interactive}, general-purpose proving strategy in \Tma.

In contrast to most other proof assistants, the interactive prover in \Tma\ is not text-based, but \emph{dialog-based}: whenever a new proof situation that cannot be handled automatically\footnote{So, there is still \emph{some} automation of very trivial tasks.} arises during the proof search, a dialog window pops up. This window displays the current proof situation, characterized by the current proof goal and the current set of assumptions, and asks the user how to proceed. He or she may now either
\begin{itemize}
 \item choose an inference rule to apply,
 \item choose a different pending proof situation where to continue with the proof search,
 \item inspect the proof \emph{so far}, in a nicely-formatted proof document,
 \item inspect the internal representation of the proof object for debugging,
 \item save the current status of the proof in an external file,
 \item adjust the configuration of the prover (maybe even switching from the interactive mode to a fully automatic one), or
 \item abort the proof attempt.
\end{itemize}
When choosing an inference rule that shall be applied (or, more precisely, \emph{tried}), the user even has the possibility to indicate the formula(s) to be considered by the rule (for instance, if one of several universally quantified assumptions is to be instantiated). Furthermore, he or she may then be asked to provide further information about the concrete application of the rule (like specifying the concrete term a formula shall be instantiated with); this, however, solely depends on the implementation of the inference rule and is thus not affected by our interactive proving strategy.

Summarizing, the interactive strategy proved to be very practical and convenient in our formal treatment of reduction rings; almost all proofs were carried out interactively. Still, it is not quite satisfactory yet: its integration into \Tma, and in particular into the \Tma-Commander window, can definitely be improved, and also a text-based interface complementing the dialog-based one is desirable.

\begin{figure}[t]
\centering
\includegraphics[width=0.8 \linewidth]{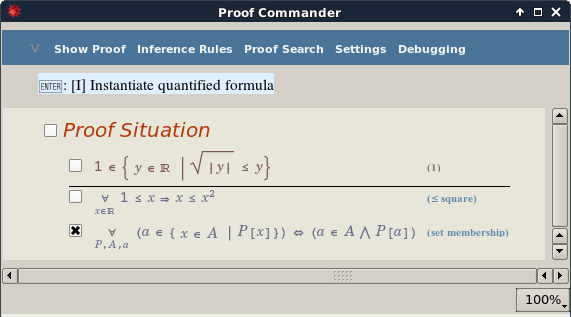}
\caption{A ``Proof Commander'' dialog window for interactive proving.}
\label{fig::Dialog}
\end{figure}

Figure~\ref{fig::Dialog} shows a screen-shot of the interactive dialog window. In the middle, the current goal (top) and the current assumptions (bottom) are displayed. Above, the inference rule to be applied next, as chosen by the user, is indicated, and the menu bar is located at the very top.

\subsection{\TA}
\label{sec::TheoryAnalyzer}

The \TA\ is a \Mma\ package that provides a collection of functions for analyzing the logical structure of \Tma\ theories and the logical dependencies of formulas on each other. If theories grow big, as in our case, it becomes more and more difficult to keep track of which formulas were used in the proofs of which other formulas, which formulas are affected when another formula is modified, and whether the order of formulas in a notebook agrees with their logical order. It is clear, however, that these questions are of utmost importance for a consistent, coherent and systematic development of a mathematical theory; after all, if a formula $\varphi$ is modified, then all of its consequences (that is, the theorems that use $\varphi$ as an assumption in their proofs) \emph{must} be re-proved, and so one needs to know what these consequences are in the first place -- and this was the main motivation for the development of the \TA.

In more concrete terms, the \TA\ works as follows:

First the user has to call a function that scans all proof files (i.\,e. external files containing information about the goal and the list of assumptions of every proved theorem) in a given list of directories. Scanning the files, the \TA\ internally constructs a directed graph whose nodes are the formulas thus found (goals \emph{and} assumptions), and whose edges resemble the logical dependency between the formulas: the graph contains a directed edge from formula $\varphi$ to formula $\psi$ iff the proof of $\psi$ uses $\varphi$ as an assumption.

As soon as this task is completed, the user can
\begin{itemize}
 \item inspect all direct or indirect assumptions of a given theorem,
 \item inspect all direct or indirect consequences of a given formula,
 \item perform an integrity check, i.\,e. check whether some theorem logically depends on itself,
 \item make sure that the order of formulas in a notebook agrees with their logical order, and
 \item ask the system to automatically draw nicely-formatted theory-dependency-graphs (as the one in Fig.~\ref{fig::ThyGraph}) and statistics diagrams (as the one in Fig.~\ref{fig::ThyStats}).
\end{itemize}

Although all the functionality listed above in the end boils down to standard graph algorithms (exhaustive search, loop detection, \ldots), it entails extensive support for the user developing a theory in \Tma. Our own experience with the formalization of reduction rings revealed that modifying formulas \emph{after} they have already been used as assumptions in proofs, re-structuring parts of theories, and even re-factoring the whole formalization happens quite frequently, and so our \TA\ will definitely aid also future theory explorations in \Tma.